\documentstyle[a4]{article}

\def\beq{\begin{equation}}
\def\eeq{\end{equation}}

\begin{document}
\title{A short derivation of Feynman formula}

\author{Alejandro Rivero \thanks {\tt rivero@sol.unizar.es} \\
{\it Dep. F\'{\i}sica Te\'orica, Universidad de Zaragoza,
                                                50009 Zaragoza, Spain}}

\maketitle
\begin{abstract}
The complex exponential weighting of Feynman formalism is seen
to happen at the classical level. 
(Finiteness of) Feynman path integral formula is suspected then
to appear as a consistency condition for the existence of
certain Dirac measures over functional spaces.
\end{abstract}

{\bf 1}. The simplest variational problems can be easily formulated in
terms of distribution theory. For instance, recall the static
problem, to find the minimum of a function and evaluate some
quantity in that minimum. We can reformulate it as: 
 Given a function $f(x)$, find a
Dirac measure $\delta_f$ concentrated in the critical points of $f$.

The answer is obviously $ \delta(f'(x))$. Its exponential form,
\beq
< \delta(f'(x)) | g(x)> = 
\int \int e^{i z f'(x)} g(x) dz dx =
\int \int \lim_{y\to x} e^{i z {f(y)-f(x)\over y-x} } g(x) dz dx
\eeq
is interesting for itself, but  we can carry it to a more amazing
shape by making the substitution $\epsilon= {y - x \over z}$. Then the
previous limit is asymptotically equivalent to 
\beq
\label{class}
<\delta_f | g> =
\int \int  \lim_{\epsilon\to 0} e^{i {1 \over \epsilon} (f(y) - f(x))}
                                              g(x)    dx  {dy \over \epsilon}
\eeq 
This is done controlling $x-y$ and  the oscillating
character of the exponential. 

Now, if $f$ has only an extremal point, we can choose 
to work with the
"halved" expression,
\beq
\label{bare.0}
<\delta_f^{1/2} | O> = \lim_{\epsilon\to 0} {1 \over \epsilon^{1/2}}
                        \int e^{i {1\over\epsilon} f(x)} O(x) dx
\eeq
from which we can recover (\ref{class}) by taking modulus square,
\beq
<\delta_f | g> = <\delta_f^{1/2} | O> <\delta_f^{1/2} | O>^*,
\eeq 
with $g(x)=O(x)O^*(x)$.
Of course, other games are possible, changing the regularization
method - i.e. the role of $\epsilon$ -, but all of them are equivalent
in the vicinity of $\epsilon= 0$.

The whole point here is that the basic structure of path integral,
the complex exponential weighting, is already present in the 
integral presentation of the most elementary variational problem,
the principle of virtual work, which we teach in 
general physics courses. 

It is intriguing to notice that Dirac guessed the exponential
weighting directly from quantum mechanics, trying to build
contact transformations \cite{dirac}, instead of uplifting
it from its classical version, as we are doing here.

{\bf 2}. To go from the zero-dimensional problem (static) to the one
dimensional (classical dynamics) we need 
to generalize (\ref{bare.0}) to spaces of functions of one
variable, time. There we can not directly assert the convergence of the
regularization, and we need to follow an indirect route, inspired in
Wilson-Kogut triangles \cite[ch. 12]{wk}. Feynman formula will appear
 as a convergence
condition: the regularizated measure has a limit if and only
if the Feynman measure over paths has a finite one.

First lets restate the question: We are given a functional $L[\phi]$ and
associated contour conditions $(\phi_0,t_0), (\phi_1,t_1)$ determining
a space $F$ of functions. The problem,
again, is to find a Dirac measure over this space $F$ concentrated in
the critical points of the functional L.
%
%
Inspired in (\ref{bare.0}), we propose as answer the
limit of the discretized functional:
\beq
\label{bare.1}
<\delta_L^{\epsilon,\epsilon'} | O[\phi] > =
 \int ... \int {1\over \epsilon^{n/2}} 
 e^{i {1\over \epsilon} \epsilon' 
       \sum_{i=0}^n  L^{\epsilon'}[x_i, {x_{i+1}-x_i \over \epsilon'} ]}
                                       O[\phi]
 (\Pi dx_i) 
\eeq
Where $\epsilon'=(t_1-t_0)/(n+1), x_0=\phi_0, x_{n+1}=\phi_1$, and we must
take both limits
$\epsilon,\epsilon' \to 0$.
Each integration in $dx_i$ is a mirror of formula (\ref{bare.0}),
concentrating in the values of $x_i$ where the function $L[x_1,...x_n]$,
takes its extremal value
keeping the rest of $\{x_j\}$ fixed.

At this point, we could directly define a proportionality constant
between $\epsilon$ and $\epsilon'$, say $\epsilon= h \epsilon'$, to join
both limits and them claim (\ref{renlimit.1}) directly.

{\bf 3}. But better we would like to try a more sophisticated process, and
ask ourselves about the convergence of the 
$\epsilon,\epsilon' \to 0$ limit. To do it, we introduce an arbitrary
scale $h$ which controls the limit process: each term
would be moved back according a transformation
\beq
\label{RG.transf}
\epsilon \to \tau_{\epsilon\over h}(\epsilon), 
\epsilon' \to \tau_{\epsilon\over h}(\epsilon'),
\eeq
whose exact form still elude us. 
Surely
we must to take in account that, in addition to $\epsilon,\epsilon'$, also
the quantities $\xi_i\equiv x_{i+1}-{x_i}$ go to zero. In fact, Feynman
proof of Schr\"odinger equation \cite{feynman} relies heavyly
in an approximation for small $\xi$. In our case, it is implied an
adjustment in $\xi$ which we can hide under the carpet of the
Lagrangian.
  
On the other side,
transformation (\ref{RG.transf}) reduces the number $n$ of points where
we fix the classical path. So, alternatively, we could try to 
build $\epsilon \to 0$  as
a discrete series of bipartitions $\epsilon_{n+1} = \epsilon_n /2$,
and then the control as a block summation back to the expression
of level $\epsilon_0=h$.

{\bf 4}. In any case,  the limit would  be convergent if and
only if the controlled series
\beq
\label{renorm.1}
<\delta_L^{h,\epsilon'} | O[\phi] > =
 \int ... \int {1\over (h \epsilon')^{n/2}} 
 e^{i {1\over h} L^{\epsilon'}_n[\phi_0,x_1,...x_n,\phi_1]} O[\phi]
 (\Pi dx_i) 
\eeq
converges (compare with the control of a Wiener measure through
the normalized brownian bridge, see e.g. \cite[sec. 3.1.2]{roep}).
 
Notice that the limit of this second series is Feynman path integral
formula,
\beq
\label{renlimit.1}
<\delta^{h,0} | O>=
   \int e^{ i {1\over h} S[\phi]} O[\phi] (d\phi),
\eeq
as searched. Note also that our indirect travel gives the normalization
factor $(h \epsilon)^\frac12$ almost directly from (\ref{bare.0}). 
Feynman \cite{feynman} prefers to get it in the course of its
approximation for small $\xi$.

Finally, note that the constant $h$ we have introduced to
control the series is arbitrary, and we can repeat the 
construction for any other value $h=\epsilon'' >0$. In this
form we get a third series
\beq
\label{dressed.1}
<\delta^{\epsilon''} | O >= 
   \int e^{i {1\over \epsilon''} S[\phi] } O[\phi] (d\phi)
\eeq 
which is also a solution of the proposed regulatization problem,
and additionally fulfills that it is invariant under the
control transformation:
\beq
\label{RGold}
\tau_\mu <\delta^h | = <\delta^{\mu h} |
\eeq

In the spirit of Wilson-Kogut transformations, we would like
to call (\ref{bare.1}),(\ref{renorm.1}),(\ref{dressed.1}) respectively
 a bare series,
a renormalized series,
and {\em the} dressed series associated to the measure being defined.

The transformation which lets invariant the dressed series "corresponds 
to"  Gell-Mann old RG transformation, its invariants
being associated to some
mean value equations.
For instance, we could "formally" manipulate eq (\ref{renlimit.1}) for
$O= \delta L / \delta \phi$ and then we get
\beq
<{\delta L \over \delta \phi}>=
\int e^{i \frac1h L[\phi]} {\delta L \over \delta \phi} d\phi=
 \int e^{i \frac1h L[\phi]} dL= \delta({1\over h})
\eeq
which is obviously invariant under the transformation.  
RG invariance in this context relates to Ehrenfest theorem.
%

{\bf 5}. As a final remark, let us to note that 
the derivation here exposed could be nicely formulated in the
framework of the tangent grupoid of the configuration space 
as defined by Connes \cite{connes.red} (see also \cite{ours,propio}
and references therein). Elements of tangent grupoid can
be "chained" as arrays of vectors $(x,x_1)(x_1,x_2)(x_2,x_3)...
(x_n,y)$, functions over them are operator kernels such
that $(a b)(x_i,x_k)= \int a(x_i,x_k) b(x_k,x_j) dx_k$.
More, Connes grupoid relates to the grupoid of paths only if
we make a scaling $(x,y,\epsilon) \to (x,y,2 \epsilon)$
after composition of arrows.

A deeper understanding of this framework
would be useful
before to try to generalize the construction to dimensions
higher than one. For dim $>1$ we would expect more exotic fixed points, and
additional entities (fermions?) are surely needed
in order to keep the properties of line integration, and stokes-like
theorems,  that
usually are codifyed inside the wedge product of differentials.
Also, known issues related to the kind of metric (Euclidean,
Minkowski...) should become more relevant.


This work has been inspired from discussions with the theoretical
teams of Zaragoza University and Costa Rica university, whose 
patience the author wants to thank. 
Partial support from project
MEC.xxx.yyy  must be acknowledged.

{\it This document is a working draft. Comments are welcome, but please
check the database \cite{find} for more recent work in the subject}


\end{document}